\documentclass[a4paper,11pt]{article}
\usepackage{pos}
\usepackage[nameinlink]{cleveref}
\usepackage{slashed} 
\usepackage{amsmath}
\usepackage{mathtools}
\usepackage{physics}

\newcommand{\imag}{\text{i}}

\title{Flavor Hierarchies in Fundamental Partial Compositeness}

\author*[a]{Florian Goertz}
\author[a]{\'Alvaro~Pastor-Guti\'errez}
\author[b,c]{Jan~M.~Pawlowski}

\affiliation[a]{Max-Planck-Institut f{\"u}r Kernphysik,\\ Saupfercheckweg 1, 69117 Heidelberg, Germany}

\affiliation[b]{Institut für Theoretische Physik, Universität Heidelberg, Philosophenweg 16, 69120 Heidelberg, Germany}
\affiliation[c]{ExtreMe Matter Institute EMMI, GSI Helmholtzzentrum für Schwerionenforschung mbH, Planckstr.\ 1, 64291 Darmstadt, Germany}

\emailAdd{florian.goertz@mpi-hd.mpg.de}
\emailAdd{alvaro.pastor@mpi-hd.mpg.de}
\emailAdd{j.pawlowski@thphys.uni-heidelberg.de}

\abstract{The idea of partial compositeness (PC) in Composite Higgs models offers an attractive means to explain the flavour hierarchies observed in nature. In this talk, predictions of a minimal UV realisation of PC, considering each Standard-Model (SM) fermion to mix linearly with a bound state consisting of a new scalar and a new fermion, are presented, taking into account the dynamical emergence of the composites. Employing the non-perturbative functional renormalisation group, the scaling of the relevant correlation functions is examined and the resulting SM-fermion mass spectrum is analysed.}

\FullConference{The European Physical Society Conference on High Energy Physics (EPS-HEP2023)\\
 21-25 August 2023\\
Hamburg, Germany\\}

\begin{document}
\maketitle

\vspace*{-11.5mm}
\section{Introduction}
\label{sec:Introduction}

\vspace{-2mm}
The concept of partial compositeness (PC) \cite{Kaplan:1991dc,Contino:2004vy,Agashe:2004rs,Contino:2006qr,Contino:2003ve} offers a promising means to address the hierarchical structure of fermion masses and mixings. Here, the mass terms are induced from linear mixings of elementary fermions of each Standard-Model (SM) flavour with composite fermionic operators ${\cal O}_B$, containing new fundamental fields that are bound together by a confining interaction. 
Below the condensation scale $\Lambda_c$, the light fermion mass eigenstates are thus a superposition of elementary SM-like fermions and composite resonances excited by the operators ${\cal O}_B$, which provide the connection to the composite Higgs (CH)~\cite{Kaplan:1983fs,Kaplan:1983sm,Dugan:1984hq} and thus to electroweak symmetry~breaking~(EWSB).

Focusing on the third-generation up-type quarks, the PC Lagrangian reads
\begin{equation}
\label{eq:PC}
{\cal L}_{\rm mix} = \frac{ \overline\lambda_q}{\Lambda_{\rm UV}^{[{\cal O}_B^q]-5/2}}\, \overline q_L {\cal O}_{B}^q + \frac{ \overline \lambda_t}{\Lambda_{\rm UV}^{[{\cal O}_B^t]-5/2}}\, \overline t_R {\cal O}_{B}^t  \, \, +{\rm h.c.}\,.
\end{equation} 
Here, $q_L$ and $t_R$ are the embeddings of the SM-like fields into irreducible representations of the global symmetry 
of the composite sector, the couplings $ \overline \lambda_{q,t}$ are dimensionless ${\cal O}(1)$ parameters at the flavor scale $ \Lambda_{\rm UV}$ and $[{\cal O}_B^{q,t}]$ are the dimensions of the composite-sector operators. 

Assuming a walking, i.e. almost conformal, behaviour between  $\Lambda_{\rm UV}$ and the IR scale $\Lambda_c$, at the latter the linear mixing couplings will read as dictated by their renormalisation group (RG) scaling
\begin{align}
    \overline \lambda_q(\Lambda_{\textrm{c}})\simeq \overline \lambda_q(\Lambda_{\textrm{UV}})\, (\Lambda_c/\Lambda_{\rm UV})^{[{\cal O}_B^q]-5/2}\,.
\end{align}
The large hierarchies between the SM Yukawa couplings are thus naturally explained by small differences in the scaling dimensions $[{\cal O}_B^q]$ which translate to exponentially large differences in the strengths of the mixings at low energies (see \cite{Contino:2010rs,Bellazzini:2014yua,Panico:2015jxa,FCDReview} for reviews).

While the initial focus in the literature was on effective low-energy descriptions of PC, more recently UV realisations have been explored \cite{Barnard:2013zea,Ferretti:2013kya,Ferretti:2014qta,Cacciapaglia:2014uja,Vecchi:2015fma,Sannino:2016sfx,Cacciapaglia:2017cdi,Agugliaro:2019wtf,Cacciapaglia:2020kgq,PhysRevLett.126.071602,Erdmenger:2020flu}, considering the fundamental degrees of freedom leading to the composites that mix with the SM-like fermions. Here, the straightforward assumption of three-fermion bound states \cite{Barnard:2013zea,Ferretti:2013kya,Ferretti:2014qta,Vecchi:2015fma} faces severe challenges since the operators' scaling dimensions need to deviate very significantly from the large canonical value of $[{\cal O}_B]_{0,\,c} =3 [{\cal F}]_c=9/2$, in order to avoid too suppressed fermion (in particular top-quark) masses. Lattice results indicate that such large anomalous dimensions are not realised~\cite{DeGrand:2015yna, Pica:2016rmv, Ayyar:2018glg, BuarqueFranzosi:2019eee}, asking for alternatives.

One such alternative, dubbed "fundamental partial compositeness"~(FPC)~\cite{Sannino:2016sfx,Cacciapaglia:2017cdi,Agugliaro:2019wtf}, assumes the composite fermions to consist of an elementary fermion ${\cal F}$ and a scalar ${\cal S}$ (see also \cite{Kagan:1991ng,Kagan:1994qg,Dobrescu:1995gz,Altmannshofer:2015esa}). The scaling dimension is then expected close to $[{\cal O}_B]_{0,\,c} =[{\cal F}]_c+[{\cal S}]_c=5/2$, which solves the issue of the too suppressed top-quark mass (${\cal S}$ could also emerge ultimately from fermions at higher scales). However, in the original works on FPC, the concrete investigation of a dynamical generation of the hierarchically light fermion masses was left open~\cite{Sannino:2016sfx,Cacciapaglia:2017cdi,Agugliaro:2019wtf,GiacomoPrivComm}. This will be explored here by analysing explicitly the possible range of anomalous dimensions of the ${\cal O}_B$ operators via the functional Renormalisation Group (fRG)~\cite{Wetterich:1992yh,Ellwanger:1993mw, Morris:1993qb}, allowing for a systematic and versatile treatment of non-perturbative effects and the study of emergent composites \cite{Gies:2001nw, Pawlowski:2005xe, Floerchinger:2009uf, Fukushima:2021ctq,Goertz:2023nii}.


\vspace{-3mm}
\section{Effective action and emergent composites}
\label{sec:EffAct}

\vspace{-3mm}
To realise FPC, fundamental scalars ${\cal S}^{\alpha, i}$ and fermions ${\cal F}^{\alpha, a}$ are introduced in the fundamental representation of the confining, techni-color (TC) like gauge group $G_{\rm TC}$, with gauge-index $\alpha$~\cite{Sannino:2016sfx,Cacciapaglia:2017cdi,Agugliaro:2019wtf,Cacciapaglia:2020kgq,FCDReview}. 
The strongly coupled TC part of the full effective action $\Gamma$, that we employ in our renormalisation-group approach, reads (with the dots indicating higher-order interactions)
\begin{align}\nonumber
\Gamma_{\rm CH}=&\,\int_x \big\{ Z_A/4 \ G_{\mu \nu} G_{\mu \nu} +\mathcal{L}_{\text{gf+ghosts}} + Z_{{\cal S}}/2\,[ \left( D_\mu {\cal S}^{i}\right)^\dagger  \left( D_\mu {\cal S}^{i}\right) +  {{\cal S}^{i}}^{\dagger} m^2_{{\cal S}}\, {\cal S}^{i}] \\[1ex] 
&\hspace{-.9cm}+ Z_{{\cal F}}\,\bar{{\cal F}}^{a} \left(\sigma_\mu D_\mu + m_{{\cal F}} \right) {\cal F}^{a} + \sqrt{Z_{\psi}\,Z_{{\cal F}}\, Z_{{\cal S}}} \ y^{i,a}_{\rm TC} \,\psi^{i,a}\, \epsilon_{ij}\, \Phi^{j}\, \epsilon_{\rm TC}\, {\cal F}^a +{\rm h.c.} + \cdots \big\}\,,
\label{eq:CHEffectiveAction}
\end{align}
where $\epsilon_{\rm TC}$ is the antisymmetric tensor of $G_{\rm TC}$ . The TC field strength and covariant derivative~read 
\vspace{-2mm}
\begin{align}
    G_{\mu \nu}= \partial_\mu A_\nu -\partial_\nu A_\mu  - \imag \, g_{\rm TC} \left[ A_\mu,  A_\nu\right]\,,\quad D_\mu =\partial_\mu - \imag \, g_{\rm TC} A_\mu \,,
\end{align}
\vspace{-6.5mm} \newline
where we keep the colour indices and generators of the TC-gauge group implicit.

Here, we focus on the minimal incarnation of $G_{\rm TC}= \text{Sp}(N_{\rm TC})$ with $N_{\rm TC}=2$ and 4 Weyl fermions~${\cal F}^{\alpha, a}, a=1,..,4$, per TC~\cite{Cacciapaglia:2017cdi,Sannino:2017utc}, resulting in the global symmetry-breaking pattern SU$(4)_{\cal F}\to \text{Sp}(4)_{\cal F}$ due to fermion condensation. The SM-like Higgs doublet thereby emerges within the resulting set of Goldstone bosons.
The TC-fermions are accordingly assumed to form weak chiral doublets ${\cal F}^{1,2}$ with vanishing hypercharge $Y=0$ and SU$(2)_L$ singlets ${\cal F}^{3,4}$~with~$Y=\mp 1/2$. 
Together with 12 complex scalar degrees of freedom per TC, forming 3 generations of colour triplets ${\cal S}_q$ with $Y=-1/6$ and colour singlets ${\cal S}_l$ with $Y=1/2$, they build the composite operators that mix with the SM quarks and leptons.
This economic realisation of FPC is accordingly called ``minimal fundamental partial compositeness" (MFPC)~\cite{Sannino:2016sfx,Cacciapaglia:2017cdi,Sannino:2017utc}.
For $m_{\cal F}\!=\!m_{\cal S}\!=\!0$, which we will assume below, the TC-fermions (scalars) exhibit a global SU$(4)_{\cal F}$ (Sp$(24)_{\cal S}$) flavour symmetry, corresponding to transformations along the index $a$ ($i$) in \labelcref{eq:CHEffectiveAction}. 
In the last line of~\labelcref{eq:CHEffectiveAction}, $y^{i,a}_{\rm TC}$ are the Yukawa couplings between the components of $\psi^{i,a}$, comprising the SM fermions embedded in the full global symmetry, and the fundamental TC-fields. Moreover, $\epsilon_{ij}$ is the anti-symmetric tensor in Sp$(24)_{\cal S}$ and the TC-scalar fields have been arranged as $\Phi=({\cal S}, - \epsilon_{\rm TC} {\cal S}^\ast)^{\rm T}$, see \cite{Cacciapaglia:2020kgq,Sannino:2017utc,Sannino:2016sfx,FCDReview,Goertz2021}.  

\begin{figure}[t]
    \centering
    \vspace{-.3cm}
    \includegraphics[width=0.51\columnwidth]{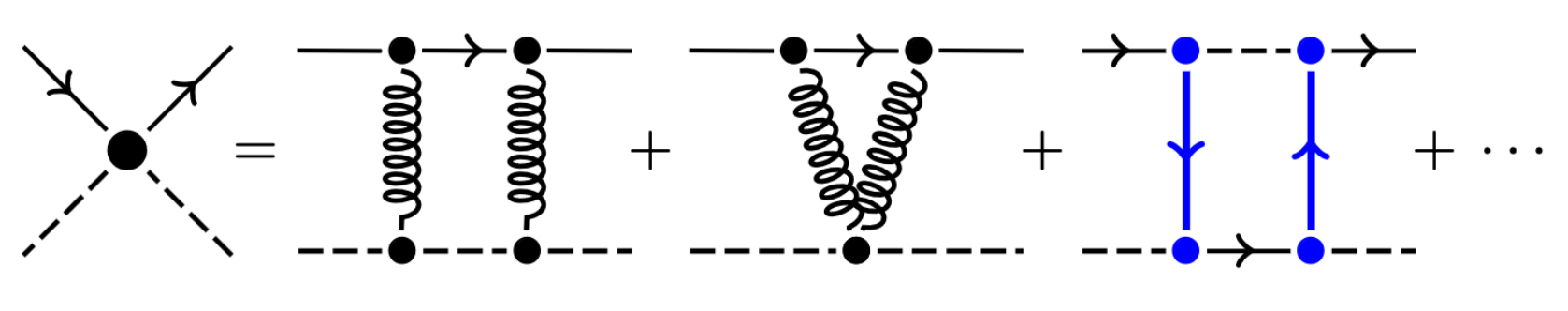}\quad 
      \includegraphics[width=0.46\columnwidth]{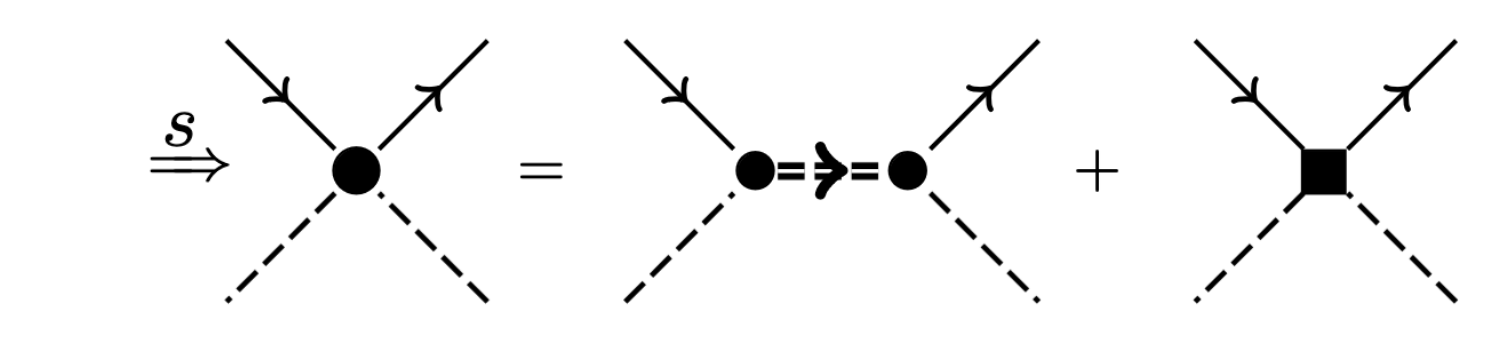}
    \vspace{-.75cm}
    \caption{Left: Generation of interaction between two TC-fermions (black arrow lines) and two TC-scalars (dashed) from the TC-gluons (curly) and SM fermions (blue arrow line) mediated box diagrams. 
    All vertices ($\bar \Gamma^{(n)}(p_1,...,p_n)$, black and blue dots) and propagators ($1/\bar \Gamma^{(2)}(p_1,p_2)$) are full correlation functions~\cite{Goertz:2023nii}. Right: Rewriting the full $s$-channel (fermion--scalar)$^2$ interaction as a composite~${\cal B}$ exchange and a remnant, containing the momentum-dependence of the resonant tensor structure and other~tensor~structures.}
    \label{Fig:FFSSgen}
\end{figure}

A relevant example of a higher order term in \labelcref{eq:CHEffectiveAction} is the two-scalar--two-fermion scattering, which is generated from the box diagrams depicted in the left panel of \Cref{Fig:FFSSgen}. 
These lead to
\vspace{-2mm}
\begin{align}
    \Gamma_{{\cal S}^2{\cal F}^2 }=\int_x 
     Z_{{\cal S}} Z_{{\cal F}} \,g_{{\cal S}{\cal F}}\,{\cal S}^\dagger {\cal S}\,\bar{{\cal F}}{\cal F} +\cdots \,, 
\label{eq:4TCinteraction}
\end{align}
\vspace{-6mm} \newline
where $\cdots$ indicate all further two-scalar--two-fermion terms.
After performing derivatives w.r.t the RG invariant fields $Z_\phi^{1/2}(p)\, \phi(p)$, as indicated by subscripts, we obtain the momentum-dependent and RG-invariant scattering coupling (neglecting Dirac, flavour, and TC tensor structures)
\begin{align} 
{\cal P}_{{\cal S}^\dagger {\cal S}\,\bar{{\cal F}}\,{\cal F}}\ \, \bar\Gamma^{(4)}_{{\cal S} {\cal S}^\dagger {\cal F}\bar{{\cal F}}}(p_1,p_2,p_3,p_4) \simeq 
g_{{\cal S}{\cal F}}(p_1,p_2,p_3),\ \ g_{{\cal S}{\cal F}}(p) \! \propto \! (r_g g_\textrm{\tiny{TC}}^4(p) + r_y y_\textrm{TC}^4(p))/p,
\label{eq:barGamma4}
\end{align}
where we dropped the momentum conservation $(2\pi)^4\delta(p_1+\cdots +p_4)$ and ${\cal P}_{{\cal S}^\dagger {\cal S}\,\bar{{\cal F}}{\cal F}}$ projects on the two-scalar--two-fermion term in \labelcref{eq:4TCinteraction}, while the relation between couplings follows from \Cref{Fig:FFSSgen}. The coupling is suppressed in the UV due to asymptotic freedom of the TC interactions, with $r_{g/y}$ being combinatorial factors of the diagrams and we focused on a symmetric point with $(p_i^2)^{1/2}=p$.

The resonant two-scalar--two-fermion scatterings may give rise to the formation of the fermionic composites ${\cal B} \sim {\cal S F}$, just as for mesons in QCD via resonant four-quark scatterings. These resonant channels can be described by the propagation of a new degree of freedom, a composite operator consisting of both fundamental TC fields, $\mathcal{O}^{a,\,i }_{\cal B} \sim \epsilon_{i\,j}\,{\cal S}^{\alpha,j}\,{\cal F}^{\alpha,a} {\cal T}$,
which may also include higher-order terms. All symmetries of the TC fields (containing the SM gauge symmetries) are encoded in the indices $a,i$, while $\mathcal{T}$ projects on the corresponding spin indices of the TC fields, keeping the construction general. Indeed, the effective four-field interaction can be exactly rewritten as the exchange of an emergent (resonant) composite of both fundamental external fields in a specific momentum channel and a residual contribution, see \Cref{Fig:FFSSgen}, right panel, which is known as \textit{dynamical hadronisation} or more generally \textit{emergent compositeness} \cite{Gies:2001nw, Pawlowski:2005xe, Floerchinger:2009uf, Fukushima:2021ctq}. The corresponding composite fields are simply introduced in the path integral via the respective current term~\cite{Pawlowski:2005xe, Fu:2019hdw, Fukushima:2021ctq},
$\exp\left\{\int_x J_{\cal B}\,{\cal B}(\varphi_f)\right\}$\,.
Here, $\varphi_f$ denote the fundamental fields being integrated in the path integral. The introduction of the composite fields via a current is just a convenient reparametrisation of the fundamental theory in terms of emergent composites and no reduction to an effective field theory.
The effective action follows from a Legendre transformation and includes the  composite's~dynamics, 
\begin{align}
\Gamma_{{\cal B}}=& \int_x \{ Z_{{\cal B}} \, \bar{{\cal B}}\, \left( \sigma_\mu \partial_\mu+ m_{{\cal B}}\right){\cal B}
+ h_{{\cal B}} \, \sqrt{Z_{{\cal B}}\,  Z_{{\cal S}}\,Z_{{\cal F}}} \left[{\cal S}\left(\bar{{\cal B}} {\cal F}  \right)+{\cal S}^\dagger \left(  \bar{{\cal F}}{\cal B} \right) \right]+\cdots\}\,,
\label{eq:Baction}
\end{align}
where we have omitted flavour and chiral indices as well as $\mathcal{T}$ and $\epsilon_{ij}$. The RG-invariant scattering coupling, neglecting again Dirac, flavour, and TC structure and momentum conservation, reads
\begin{align}\label{eq:Gamma3BSF}
\bar \Gamma^{(3)}_{{\cal S}{\cal F}\bar{{\cal B}}}(p_1,p_2,p_3) \simeq h_{\cal B}(p_1,p_2)\,.
\end{align}

Importantly, \labelcref{eq:Baction} encodes the diagrammatic relation of \Cref{Fig:FFSSgen} via the equation of motion (EOM) of the composite field ${\cal B}$. For $s$-channel configurations $s\!=\!(p_1+p_2)^2$ this relates \labelcref{eq:Baction}~to~\labelcref{eq:4TCinteraction}~with 
\begin{align}
\left. \Gamma_{{\cal B}}\right|_{ {\cal B}_{\textrm{\tiny{EoM}}}} = \Gamma_{{\cal S}^2{\cal F}^2 }\ ,\quad 
 (h^2_{\cal B}(s)\,m_{\cal B}(s))/(s+m_{\cal B}^2(s))\propto g_{{\cal S}{\cal F}}(s) \,, 
 \label{eq:GBGSF}
\end{align}
and $m_{\cal B}(s)\propto \sqrt{s}$.
It follows from \labelcref{eq:barGamma4} that $h^2_{\cal B}(s) m_{\cal B}(s)/(s+m_{\cal B}^2(s))\propto g_\textrm{\tiny{TC}}^4(s)/\sqrt{s}$ (assuming the TC gauge interactions to dominate), and 
given the two-scalar--two-fermion $s$-channel scattering becomes resonant in the IR, this is well described in terms of the ${\cal B}$-exchange. 
We note that the full effective action in the presence of the composites includes \labelcref{eq:CHEffectiveAction,eq:Baction} and that the approach at hand allows for a global description in terms of the dominant degrees of freedom, linking the low-energy interactions with those in the fundamental high energy theory, which is not possible in standard effective theory approaches to strongly coupled theories.

After introducing the dynamical composites in the effective action, also the linear mixing terms
\begin{equation}
\Gamma_{\rm mix}=\int_x \Bigl\{\,\lambda^L_t  Z^{1/2}_q Z^{1/2}_{{\cal B}} \ \bar{q}_L {\cal B}^q_R +\lambda^R_t Z^{1/2}_t Z^{1/2}_{{\cal B}} \ \bar{t}_R {\cal B}^t_L
+\text{ h.c.}\Bigr\}\,,
\label{eq:qBaction} 
\end{equation}
allowed by the symmetries, will be generated - responsible for the SM fermion masses in FPC. The scattering vertices of the fundamental fields are obtained by taking derivatives of the fundamental effective action, or, more conveniently, of the action including the composite (Eqs.~\labelcref{eq:CHEffectiveAction,eq:Baction,eq:qBaction}) on the EOM, w.r.t these fields (see \labelcref{eq:GBGSF}). Via the relation in \Cref{Fig:FFSSgen}, the latter approach avoids the non-practical evaluation of higher terms in the fundamental fields. The EOM solution reads
\begin{align}
    {\cal B}^{\textrm{\tiny{EoM}}}=  \sqrt{Z_{\cal S} Z_{\cal F}/Z_{\cal B}} \ h^2_{\cal B}/(\sigma_\mu\partial_\mu+m_{\cal B}) \, {\cal S} {\cal F} +\sqrt{Z_{f}\!/Z_{\cal B}}\ \lambda_f\, f\,,
\label{eq:BEoM}
\end{align}
with $f$ the respective SM fermion. Performing $\psi,{\cal S},{\cal F}$-derivatives, we obtain the relation
\begin{align}
  y^{f}_{\rm TC} = y^{f}_{\rm c, TC} -\lambda_f \,(h^2_{\cal B}\,  m_{\cal B})/(p^2+ m_{\cal B}^2)
\label{eq:ycy}
\end{align}
between the Yukawa coupling in the fundamental effective action~\labelcref{eq:CHEffectiveAction} and the (numerically different) parameters of the full action including the composite, featuring an additional contribution from~\labelcref{eq:qBaction}.

We now discuss the anomalous momentum scaling of the dimensionless $\overline \lambda_f= \lambda_f/k$, with $k$ the average momentum, being the cutoff in the fRG approach. The anomalous dimension~$\gamma_{\lambda_f}$~is~defined by
$\partial_t \overline \lambda_f= \gamma_{\lambda_f} \, \overline \lambda_f$, where $\partial_t \equiv k \, \partial_k$, and the anomalous scaling of the wave functions $Z_\phi$ by $\partial_t Z_\phi =  \gamma_\phi \,Z_\phi$.  
For the mixing term~\labelcref{eq:qBaction}, the full momentum dimension, defined with~its~$k$~scaling,~reads
$[\overline \lambda_f \,k \,Z_{f}^{1/2}  Z_{\cal B}^{1/2}] =\gamma_{f{\cal B}}$. From this, we can relate the value
of the mass mixing parameter 
at a measurable IR scale $k=\Lambda_\textrm{c}$ to a UV value $\bar\lambda_f$ at $k=\Lambda_\textrm{UV}$ via its momentum scaling as
\begin{align}\label{eq:scalinglambdaLambdaUVLambdac}
\overline \lambda_f (\Lambda_{c})= \overline \lambda_f(\Lambda_\textrm{UV}) \left(\Lambda_\textrm{c}/\Lambda_\textrm{UV}\right)^{\gamma_{\lambda_f} }\,, \quad \text{with } 
\gamma^{\ }_{\lambda_f} \! \equiv \! \left[\overline\lambda_f\right] = \gamma_{f\cal B} -1  -\gamma_f/2 -\gamma_{\cal B}/2\,
\end{align}
enclosing all non-perturbative information on the composite and its dynamics. 

With $\gamma_{f}$ being negligible, the magnitude and sign of $\gamma_{\cal B}$ and $\gamma_{f \cal B}$ will lead to a more or less enhanced scaling of $\overline \lambda_f$. This determines the SM-like fermion masses due to mixing with the resonances (after integrating out the latter, see, e.g.~\cite{Goertz2021,Goertz:2021xlx,Panico:2015jxa}) as
\begin{align}\label{eq:m}
m_f \sim\! \frac{v}{\sqrt{2}} \frac{\overline \lambda_f^L (\Lambda_{c}) \overline \lambda_f^R (\Lambda_{c})}{g_{\cal B}(\Lambda_{\textrm{c}})} = \frac{\mathcal{Y}_f v}{\sqrt{2}}\left( \frac{\Lambda_{\textrm{c}}}{\Lambda_{\textrm{UV}}}\right)^{\!(\gamma^{\ }_{\lambda^{L}_f}+\gamma^{\ }_{\lambda^{R}_f})}
\,,\quad \text{with } \mathcal{Y}_f= \frac{\overline{\lambda}^{L}_f (\Lambda_{\textrm{UV}}) \, \overline{\lambda}^{R}_f (\Lambda_{\textrm{UV}})}{g_{\cal B}(\Lambda_{\textrm{c}})}
\end{align} 
gathering all ${\cal O}(1)$ quantities (obeying the naturalness criterion), where $g_{\cal B}$ is the coupling strength of the resonances.
The hierarchical pattern of SM fermion masses (and mixings) will now arise from small differences in the anomalous dimensions $\gamma_{\lambda}$ via RG flow between the largely separated scales $\Lambda_{\textrm{c}}\sim 10\,{\rm TeV} \ll \Lambda_{\textrm{UV}}$. We stress that
$\gamma_{\lambda_f}>0$ is necessary in the FPC framework in order to generate the masses of the light SM fermions, which we will check below.

\vspace{-3mm}
\section{Anomalous scaling and flavour hierarchies}\label{sec:fRG+MassHierachies}

\vspace{-3mm}
In this section, we compute the anomalous dimensions of the composite sector in the non-perturbative fRG approach~\cite{Wetterich:1991be,Wetterich:1992yh,Morris:1993qb,Ellwanger:1993mw} with its extension to treat emergent compositeness \cite{Gies:2001nw, Pawlowski:2005xe,Floerchinger:2009uf, Fu:2019hdw, Fukushima:2021ctq}, widely employed in the areas of condensed matter and QCD. This approach implements the Wilsonian idea of progressive integration of momentum shells by suppressing quantum fluctuations of momenta $p$ below an IR cutoff scale~$k$ via a regulator $R^{(\phi)}_k$\!.\ This leads to a cutoff dependent effective action $\Gamma_{k}[\phi]$, with its flow determined by the so-called Wetterich equation,~see~\cite{Dupuis:2020fhh}~for~a~review.

\begin{figure}
\centering
\includegraphics[width=0.6\columnwidth]{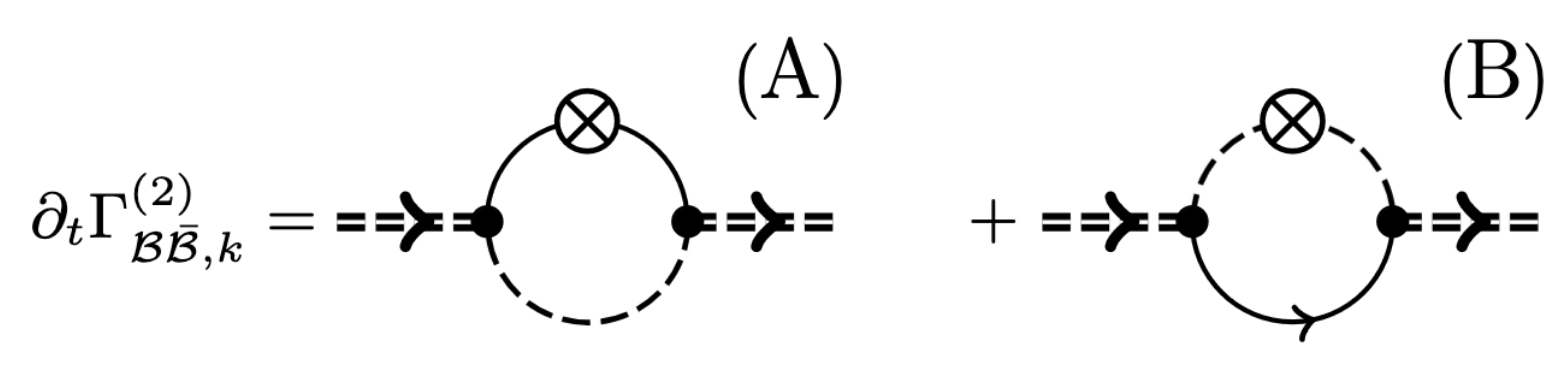}
\vspace{-0.4cm}
\caption{Diagrammatic flow of the composite's two-point function, necessary for the computation of the composite's anomalous dimension \labelcref{eq:gamB}. All lines correspond to regulated propagators. Crossed circles depict the insertion of the derivative of the fRG regulator $\partial_t R^{(\phi)}_k$. 
}
\label{fig:twopointB}
\end{figure}

First, it turns out that, due to the tensor structure in the respective diagrams, the anomalous dimension $\gamma_{f{\cal B}}$ vanishes for $m_{\cal F}=0$~\cite{Goertz:2023nii}, such that in \labelcref{eq:scalinglambdaLambdaUVLambdac} only $\gamma_{\cal B}$ remains to be calculated explicitly. For this, we observe that the momentum scale variation of the wave-function renormalisation can be obtained from the composite's two-point function 
\begin{align}
\Gamma^{\left(2\right) }_{{\cal B}\bar{\cal B},k}= \frac{\delta \Gamma_k}{\delta {\cal B}(p)\delta {\bar{ \cal B}}(-p)}=  \imag Z_{{\cal B}}\, \sigma_\mu\, p_\mu\,, \ \ \text{as \,} 
\gamma_{\cal B}\!=\!\frac{\partial_t  Z_{\cal B}}{  Z_{\cal B}}\!=\! \frac{-\imag  \, \, \partial_{p^2}\left(\, \sigma_\mu\, p_\mu\,\, \partial_t \Gamma^{\left(2\right) }_{{\cal B}\bar{\cal B},k} \right)}{  Z_{\cal B}\,\tr \, [\sigma_\mu \sigma_\nu ]}\Big|_{p=0} \,. 
\label{eq:gamB}
\end{align}
The flow $\partial_t \Gamma^{(2)}_{{\cal B}\bar{\cal B},k}$ is now derived by solving the (one-loop exact) Wetterich equation, with its diagrammatic form depicted in \Cref{fig:twopointB}. Explicit expressions and more details are provided in~\cite{Goertz:2023nii}.
Diagram (A) vanishes for $p=0$ and thus the anomalous dimension is driven by diagram (B). We obtain the analytic result
\begin{align}\label{eq:gammaB}
\gamma_{{\cal B}}=& \,-\frac{ h^2_{{\cal B}}}{16\pi^2}\frac{N_{\rm TC}}{2} \tr[\mathcal{T}^2]\left(1+\frac{\gamma_{{\cal S}}}{5}\right)\ <0\,,
\end{align}
where $\gamma_{\cal S}$ is the anomalous dimension of the fundamental scalar TC-field, whose subleading effect can be neglected for the time being (as will a potential $\tr[\mathcal{T}^2] \neq 1$), and  $N_{\rm TC}$ is the dimension of the fundamental representation of the SU($N_{\rm TC})$-gauge group. 

\begin{figure*}[t!]
\centering
    \includegraphics[width=.85\textwidth]{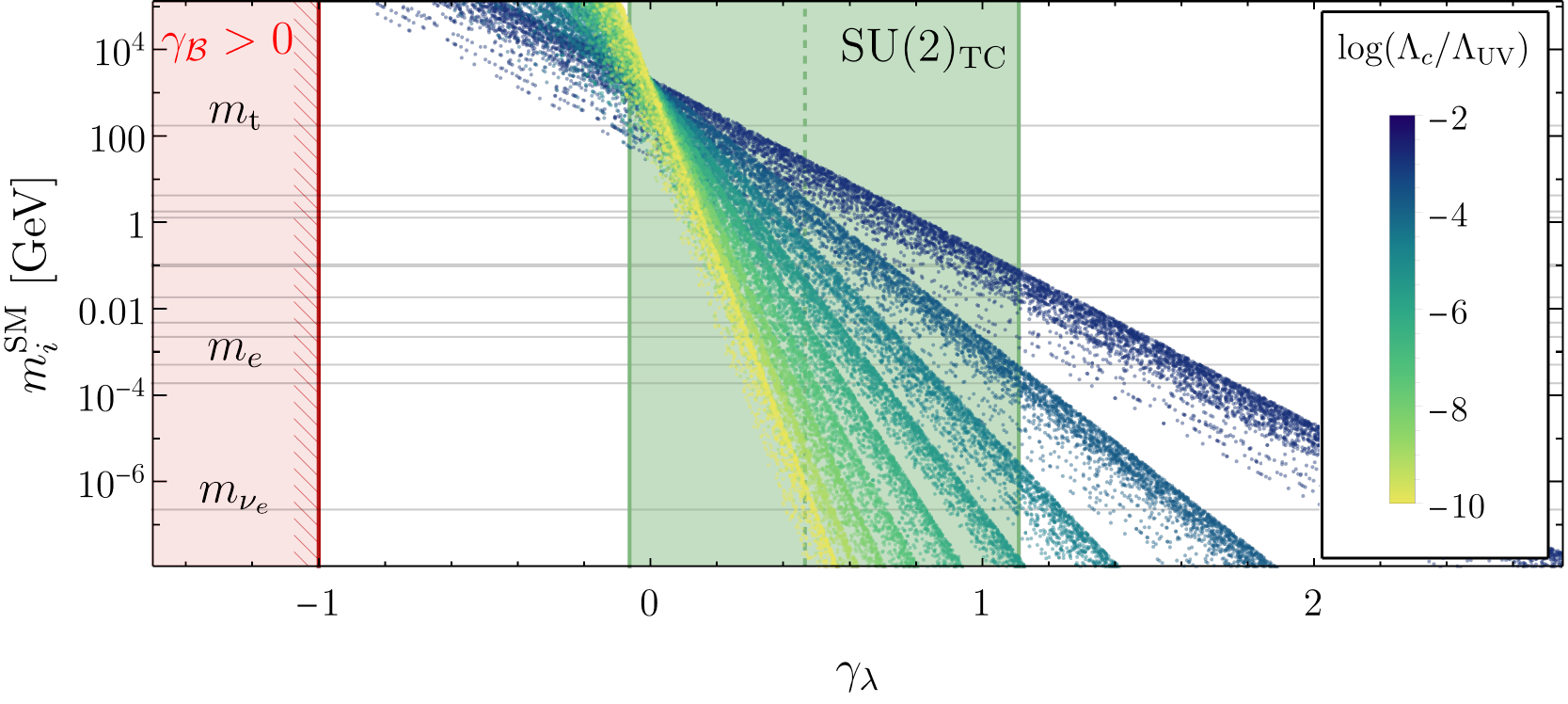}  \ \ \ \
    \vspace{-0.3cm}
    \caption{SM fermion masses as a function of the anomalous dimension $\gamma_\lambda$. The scans correspond to ${\cal Y}_f\in  [0.1,4\pi]$, $\gamma_\lambda \in [-1,3]$, and the size of the walking regime is varied within $\Delta\Lambda\equiv\log\left(\Lambda_{\textrm{c}}/\Lambda_{\textrm{UV}}\right)\in [-2,-10]$, as indicated by colours. While the red-shaded region is strictly excluded, the green region indicates the  ballpark of $\gamma_\lambda$ in the MFPC scenario with two additional Dirac fermions in \labelcref{eq:estimategammalambda}, see text~for~details.}
	\label{Fig:SMgammaNP}
\end{figure*}

In \Cref{Fig:SMgammaNP} we illustrate the SM fermion masses from \labelcref{eq:m} as a function of the anomalous dimension, where we assume  $\gamma_{\lambda^{L}_f}=\gamma_{\lambda^{R}_f}\equiv\gamma_\lambda \ (\approx -1\!-\! \gamma_{\cal B}/2)$. Scans in ${\cal Y}_f\subset [0.1,4 \pi]$ for nine different walking regimes of sizes $\Delta\Lambda \subset[-2,-10]$ are shown in different colours. 
While for the ${\cal O}$(1) top Yukawa $\gamma_\lambda\sim 0$ is needed, the lightest fermions require $\gamma_\lambda\sim 0.5-2.0$, depending on the walking regime.  
Finally, in order to provide a quantitative prediction for the anomalous dimension in models of MFPC, we employ~\labelcref{eq:GBGSF} to estimate the composite sector parameters from the fundamental TC couplings in the walking regime. Here, we extend the minimal MFPC matter content with additional Dirac fermions in order to allow for a walking regime, which is not present for the minimal content~\cite{Dietrich:2006cm,Pica:2010xq}, and employ the 4-loop $\overline{\textrm{MS}}$ results~\cite{vanRitbergen:1998pn,Vermaseren:2000nd,Dietrich:2006cm,Pica:2010xq,Fukano:2010yv} in our numerics.  Our result for 1, 2 and 3 additional Dirac fermions reads
\begin{align}\label{eq:estimategammalambda}
    \gamma_\lambda \sim -1 - \frac{1}{2}\left[-\frac{N_{\rm TC}}{32 \pi^2}\left(\frac{(1+\bar{m}_{\cal B}^2)\, r_g\, (g^*_{\rm TC})^4}{\bar{m}_{\cal B}}\right)\right] \sim \{4.74,\,0.47,\,-0.42 \}\,,
\end{align}
where we considered $\bar{m}_{\cal B} \equiv m_{\cal B}/\sqrt s =1$ and $r_g\approx 1$, see \cite{Goertz:2023nii} for more details.
The estimate for MFPC with two additional Dirac fermions is presented as a green vertical dashed line, while the green shaded region encloses a 20$\%$ variation in $h_{\cal B}$.

\vspace{-3mm}
\section{Conclusions}\label{sec:Conc}
\vspace{-3mm}

We have explored the generation of the SM fermion mass hierarchies in FPC via the non-perturbative fRG. This provides a novel application of functional methods to new physics scenarios involving strong dynamics.
The dynamical treatment of the emergence of resonances allows us to investigate the theory in a global manner, taking into account fundamental and composite degrees of freedom simultaneously as well as their interplay. We can thus access the properties of the composites, such as their couplings, from the parameters of the fundamental effective action. In particular, we derived the anomalous scaling of the linear mixing couplings from the momentum scaling of the 2-point functions in the effective action via the fRG. Finally, we presented an estimate for the mass spectrum in the MFPC scenario with two additional Dirac fermions coupled to the TC-gauge group, which confirms the possibility to address the flavor puzzle in FPC.

\section*{Acknowledgements}
\vspace{-4mm}

FG would like to thank the organisers of EPS-HEP2023 for the opportunity to present this work at the conference.

\vspace{2mm}

\small
\bibliographystyle{JHEP}
   \bibliography{references}

\end{document}